\documentclass[12pt]{article}

\usepackage{scicite}
\usepackage{siunitx}
\usepackage{todonotes}

\usepackage{graphicx}

\usepackage{times}

\newenvironment{sciabstract}{%
\begin{quote} \bf}
{\end{quote}}

\title{Solute effects in confined freezing} 

\author
{Felix Ginot$^{1,2}$, Th\'{e}o Lenavetier$^1$, Dmytro Dedovets$^{1,3}$, Sylvain Deville$^{1\ast}$\\
\\
\normalsize{$^{1}$LSFC, UMR 3080 CNRS/Saint-Gobain CREE, Saint-Gobain Research Provence, Cavaillon, France}\\
\\
\normalsize{$^{2}$ now with: Fachbereich Physik, Universit\"at Konstanz, Universit\"atsstrasse 10, 78464 Konstanz, Germany}\\
\\
\normalsize{$^{3}$ now with: LOF, UMR 5258 CNRS/Solvay, Univ. Bordeaux, Pessac, France}\\
\\
\normalsize{$^\ast$To whom correspondence should be addressed; E-mail:  sylvain.deville@saint-gobain.com.}
}

\date{}

\usepackage{lineno}

\begin{document} 

\baselineskip24pt

\maketitle 


\begin{sciabstract}
The presence of liquid water in frozen media impacts the strength of soils, the growth of frost heave, plant life and microbial activities, or the durability of infrastructures in cold regions. If the effect of confinement on freezing is well known, water is never pure and solutes depressing the freezing point are naturally found. Moreover, the combination of confinement and solute is poorly understood. We imaged the freezing dynamics of water in a model porous medium with various salt (KCl) concentrations. We showed that the freezing front, initially heterogeneous due to confinement, drives salt enrichment in the remaining liquid, further depressing its freezing point. Confinement and solute have a synergistic effect that results in much larger mushy layers and greater freezing point depression. These results should help understand the distribution of water in frozen porous media, solute precipitation and redistribution in soils, and cryo-tolerance of construction materials and organisms.
\end{sciabstract}


As temperature is lowered below $\SI{0}{\celsius}$, some water in porous media such as soils or construction materials remains unfrozen, impacting their physical and mechanical properties~\cite{Coussy2005}. The motion of liquid water in soils is responsible for frost heave, which damages infrastructures such as roads, train tracks, pipelines and buildings in cold regions~\cite{Rempel2010}, and could explain the fluidized transport of sediments on Mars~\cite{Oyarzun2003}. Freezing is also used as a confinement method for pollution in soils, decreasing the permeability of soils~\cite{Iskandar1986}. Understanding the freezing of water in confined media is thus essential in many domains.

Water in a confined environment at equilibrium does not freeze readily at $\SI{0}{\celsius}$ because of adsorption and interfacial curvature effects~\cite{Wettlaufer1999a}. While such effects are negligible above the microscale ($\SI{}{\micro\m}$), the freezing point depression can be as low as several tens of degrees in nm- or \AA-size pores~\cite{Christenson2001}. Water is however rarely pure and impurities such as ions, or organic matter are usually present. The ions ($\textrm{Ca}^{2+}$, $\textrm{K}^{+}$, $\textrm{Na}^{+}$) commonly encountered in soils or constructions materials (because of de-icing) are well-known for depressing the freezing point of water. As ice is formed pure, all solutes are expelled and concentrated near the growing ice. This not only further depresses the freezing point, but also creates gradients of solute. The concentration of solute may result in precipitation and impact plant life and microbial activities~\cite{Hayashi2013}, while the presence of liquid pockets and films weakens the strength of the soil~\cite{Coussy2005} and provide pathways for water motion and solute redistribution through the soils or sea ice~\cite{Lake1970}. Although the migration of solutes has been investigated in soil physics~\cite{Henry1988} and  microbiology~\cite{Kim2018a}, observations and understanding of the synergistic effects of confinement and solute at a local scale are still missing. To date, studies have mostly focused on measuring the relationship between the freezing point and the solute content~\cite{Banin1974}, assessing the macroscopic redistribution of solutes in soils during~\cite{Watanabe2001b} or after freeze/thaw cycles~\cite{Konrad1990}, its crystallization in pores~\cite{Scherer1999}, and quantifying the amount of unfrozen water in soils by NMR or calorimetry~\cite{Watanabe2002b,Watanabe2009,Bing2011a}.

Here we show by \emph{in-situ}, local observations, that the combined effect of confinement and solute during freezing is more than the sum of individual effects, and that this synergy is explained by the interplay between the local thermodynamics of freezing and solute rejection and concentration at the pore scale.

\begin{figure}[h]
\begin{center}
\includegraphics[width=12cm]{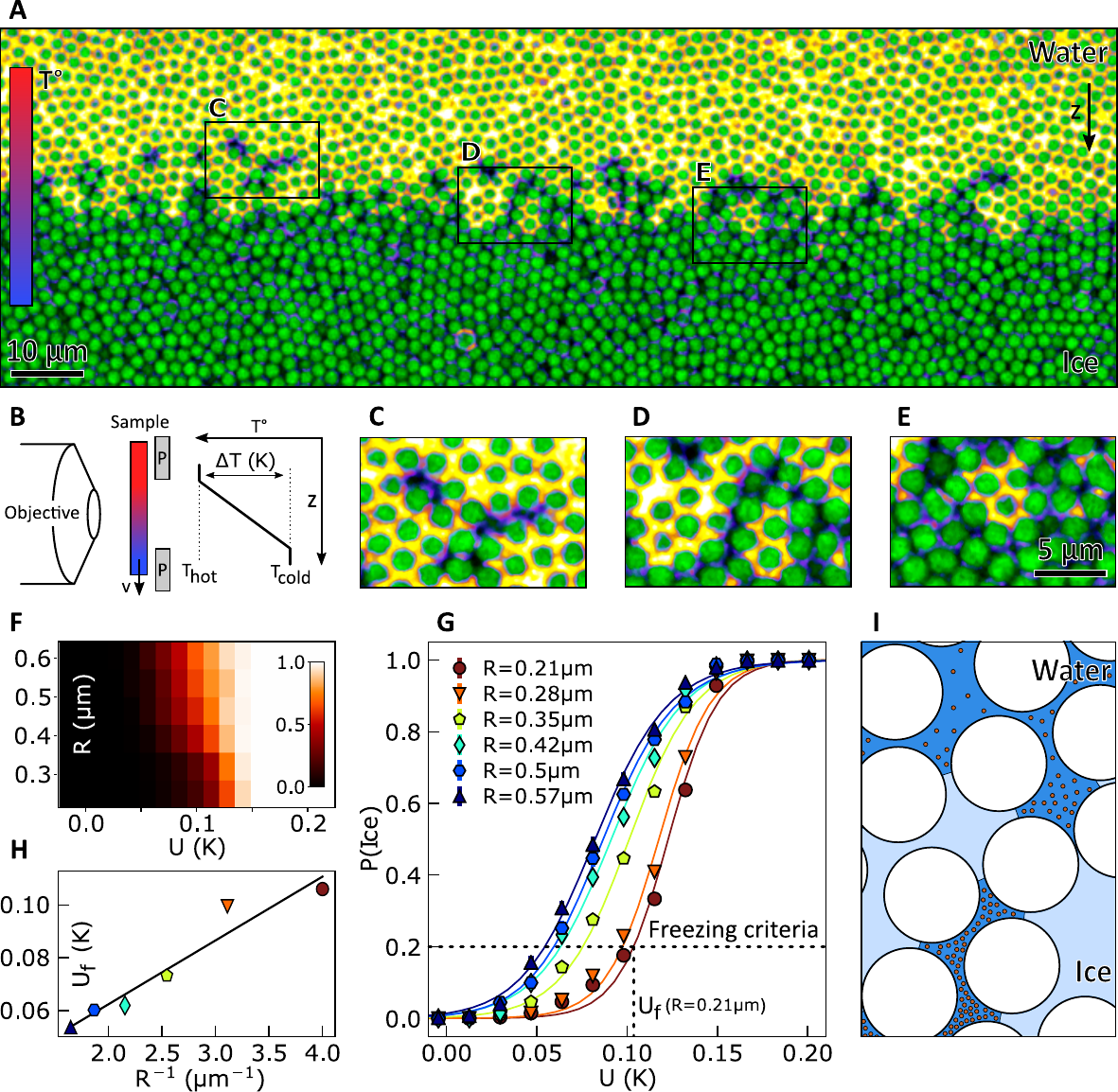}
\caption{\textbf{Confined freezing without solute.} \textbf{A.} Typical picture of freezing into a porous media made of randomly packed fluorescent particles. The porous media is shown in green, and liquid water in yellow. \textbf{B.} Sketch of the freezing setup. Two Peltiers modules provide a controlled temperature gradient. \textbf{C.} Close-up view of an isolated region where ice first appears. \textbf{D.} Close-up view of the convoluted interface between ice and water. \textbf{E.} Close-up view of an isolated liquid region within a frozen shell. \textbf{F.} 2D probability for water to be frozen in a pore of size $\textrm{R}$ and for undercooling $\textrm{U}$. \textbf{G.} Probability to find frozen water for increasing pore size $\textrm{R}$ as a function of undercooling $\textrm{U}$. \textbf{H.} Freezing undercooling $\textrm{U}_\textrm{f}$ as a function of the inverse of pore size $\textrm{R}^{-1}$.  \textbf{I.} Schematic representation of confined freezing and the concentration of the excluded solute in the remaining liquid. Red dots represent solute in solution.}
\label{fig:Confinement}
\end{center}
\end{figure}

We investigate confined freezing by confocal microscopy~\cite{Dedovets2017a}, translating a Hele-Shaw cell enclosing a porous media made of randomly packed monodispersed polymeric particles. The sample is moved at a constant velocity $V=\SI{2}{\micro\metre\per\second}$ through a constant temperature gradient $\nabla T = \SI{10}{\K\per\milli\metre}$ (Fig.~\ref{fig:Confinement}B). We record the freezing process with two photodetectors: one to track the porous media made of fluorescent particles, the other to detect a fluorescent dye (sulforhodamine 101) added to water and expelled by the growing ice. We can therefore follow \emph{in-situ} the freezing process through the confined media.

A typical confocal image is shown in Fig.~\ref{fig:Confinement}A. The propagation of freezing (movie~S1) does not occur homogeneously. Not only does the freezing front not follow isotherms---its boundary is convoluted---but isolated frozen regions are present. Frozen and liquid regions can coexist at the same temperature (Fig.~\ref{fig:Confinement}D). Isolated regions that freeze at a higher temperature than the rest of the porous matrix seem to correspond to larger pores (Fig.~\ref{fig:Confinement}C). Conversely, the densely packed regions (small pores) freeze last (Fig.~\ref{fig:Confinement}E). 

In absence of solute, confinement alone can explain this behavior. The freezing-point depression induced by confinement origins from the Gibbs-Thomson effect. In a porous media, the freezing point depression $\Delta T_f$ writes:

\begin{equation}
	\Delta T_f = T_m - T_f =  \frac{k_g T_m \gamma_{iw}}{\textrm{R} \rho_i  \Delta H_f },
\label{eq:GT}
\end{equation}

where $T_m=\SI{273}{\kelvin}$, $\rho_i = \SI{0.92}{\g\per\cubic\centi\m}$, $\gamma_{iw} = \SI{29}{\milli\J\per\square\m}$, $\Delta H_f = \SI{333}{\J\per\g}$, and $\textrm{R}$; are the bulk melting temperature, the ice density, the ice/water surface tension, the bulk enthalpy of fusion, and the pore radius respectively. $k_g$ is a geometric prefactor that depends on the pore structure.

Assuming a linear temperature gradient, we compute a relative local temperature $T_l$. Then we define an undercooling value $\textrm{U} = T_m - T_l$, sets so $\textrm{U} = \SI{0}{\K}$ at the beginning of the freezing process. Note that by convention when the undercooling $\textrm{U}$ increases, the local temperature $T_l$ decreases. Each pixel in the image is thus assigned a state (liquid or solid), a local undercooling $\textrm{U}$, and a local pore size $\textrm{R}$ (see SOM for details). We can therefore quantitatively follow \textit{in situ} the freezing of water into the pores.

The probability for water to be frozen for different pore sizes $\textrm{R}$, and increasing undercooling $\textrm{U}$, is plotted in Fig.~\ref{fig:Confinement}G. As expected, ice is more likely found in bigger pores, and at higher undercooling (lower temperature). Between the largest ($\textrm{R}=\SI{0.57}{\micro\m}$) and the smallest ($\textrm{R}=\SI{0.21}{\micro\m}$) pores, we find that confinement results in a $\sim \SI{0.05}{\K}$ freezing temperature depression.
Note that in the absence of kinetics---and perfectly defined pore size---we expect a sharp freezing transition. Here we find that this transition (that we attribute to kinetics) occurs over $\sim \SI{0.05}{\K}$, similar to the effect of confinement. For a given pore size $\textrm{R}$, we define the freezing undercooling $\textrm{U}_\textrm{f} = \textrm{U}(\textrm{P(Ice)}=0.2)$ when $20\%$ of the  water is frozen. We verify the Gibbs-Thomson relationship by plotting $\textrm{U}_\textrm{f}$ as a function of the inverse of pore size $\textrm{R}^{-1}$ (Fig.~\ref{fig:Confinement}H). As expected from Eq.~\ref{eq:GT}, a linear relationship is observed. We also find $k_g=2.2$, close to the theoretical value for perfectly cylindrical pores ($k_g^{cyl} = 2$). Such quantitative agreement suggests that the freezing point depression measured is due to confinement only, and thus that we are close to equilibrium.

\begin{figure}
\includegraphics[width=18cm]{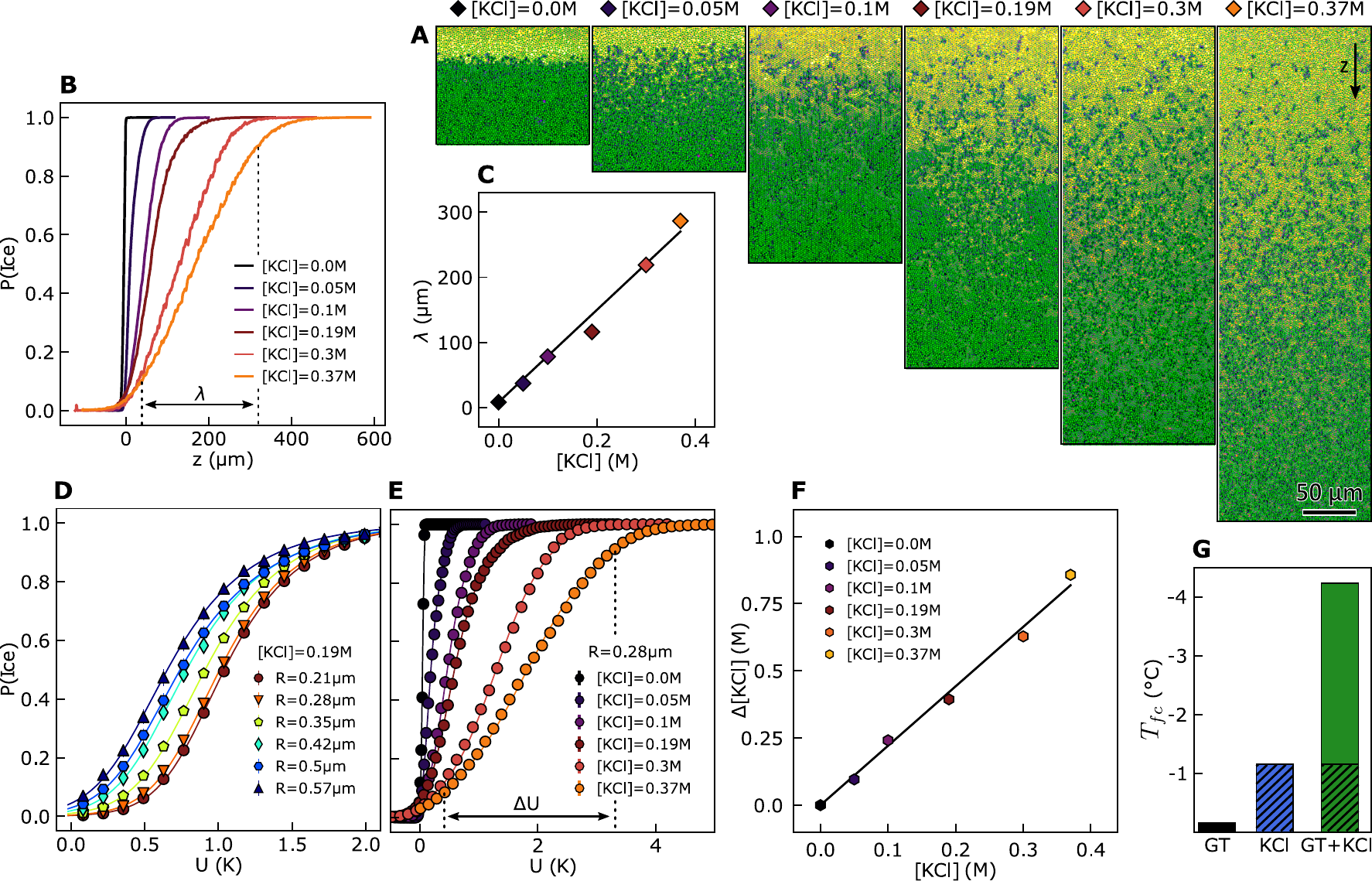}
\caption{\textbf{Confined freezing with solute (KCl)}. \textbf{A.} Confocal images of the mushy layer for increasing salt concentrations. \textbf{B.} Probability to find ice as a function $\textrm{z}$ position, and for increasing salt concentration $[\textrm{KCl}]$ from $0~\textrm{M}$ to $0.37~\textrm{M}$. \textbf{C.} Extension of the mushy layer thickness $\lambda$ increasing linearly with $[\textrm{KCl}]$ concentrations. \textbf{D.} Probability to find ice as a function of undercooling $\textrm{U}$ and increasing pore size ($[\textrm{KCl}] = 0.19~\textrm{M}$). \textbf{E.} Probability to find ice as a function of undercooling $\textrm{U}$ for increasing $[\textrm{KCl}]$ concentration ($\textrm{R} = \SI{0.28}{\micro\m}$).  \textbf{F.} $\Delta [\textrm{KCl}]$ differential of solute concentration between the start and the end of the freezing front. A three-fold increase of salt concentration is expected. \textbf{G.} Freezing completion temperature $T_{fc}$ (temperature at which 90\% of the water is frozen) for confinement alone (Gibbs-Thomson effect), solute alone, and confinement and solute ($[\textrm{KCl}]=0.37~\textrm{M}$). The hatched areas correspond to the pure solute contribution, and the plain areas to the solute/confinement synergistic effect (see SOM).}
\label{fig:Solute}
\end{figure}

Having validated the approach to study confined freezing, we now look at the impact of solute. As solutes are generally expelled from ice during freezing, we can expect them to accumulate within liquid pockets surrounded by ice shells (Fig.~\ref{fig:Confinement}I), affecting the freezing process. We use KCl as a generic solute, at concentrations commonly encountered in soils~\cite{Parida2005}. Typical pictures of freezing into the porous media with increasing $[\textrm{KCl}]$, from $0~\textrm{M}$ to $0.37~\textrm{M}$, are shown in Fig.~\ref{fig:Solute}A. The probability to find ice as a function of the $z$ position, decreases with the solute concentration (Fig.~\ref{fig:Solute}B). As a result, the thickness $\lambda$ of the mushy layer, where ice and water coexist (see SOM), increases linearly with concentration (Fig.~\ref{fig:Solute}C). For the highest concentrations ($[\textrm{KCl}]>0.19~\textrm{M}$) the freezing front becomes ill-defined, and isolated liquid pockets (and frozen islands) can be observed (Fig~S3). Most salts colligatively depress the freezing point of water. We can thus take into account the solute (see SOM) and rewrite the Gibbs-Thomson equation as:

\begin{equation}
	\Delta T_f = \frac{k_g T_m \gamma_{iw}}{\textrm{R} \rho_i  \Delta H_f} + \frac{2 k_b N_a T_m^2}{\Delta H_f \rho_w} [\mathrm{KCl}],
\label{eq:GT_Solute}
\end{equation}

with $k_b$ the Boltzman constant, $N_a$ the Avogadro number, and $\rho_w=\SI{1}{\kg\per\l}$ the density of water.

We reiterate the analysis developed in the solute-free experiments, and get a deeper look at the freezing process in the presence of salt. We show in Fig.~\ref{fig:Solute}D  the probability to find frozen water at a given undercooling $\textrm{U}$, and different pore sizes $\textrm{R}$, with $[\textrm{KCl}] = 0.19~\textrm{M}$. Similar to the solute-free experiments, small pores freeze later (at lower temperature) than big ones. However, the freezing process occurs over a much broader range of temperature. For instance, at $[\textrm{KCl}] = 0.19~\textrm{M}$, pores of size $\textrm{R}=\SI{0.21}{\micro\m}$ freeze approximately $\SI{1}{\K}$ after pores of size $\textrm{R}=\SI{0.57}{\micro\m}$. This temperature gap is more than one order of magnitude larger than for solute-free experiments ($\sim \SI{0.05}{\K}$), and therefore cannot be explained by confinement only.

We repeat the analysis for different salt concentrations to separate the solute and confinement effects. We show in Fig.~\ref{fig:Solute}E the probability for water to be frozen at a given undercooling $\textrm{U}$, for a fixed pore size $\textrm{R}=\SI{0.28}{\micro\m}$, and increasing salt concentration $[\textrm{KCl}]$ ($0~\textrm{M}$ to $0.37~\textrm{M}$). As expected, solute decreases the probability to find ice for a given undercooling value. Nonetheless, we notice that even for a fixed pore size and initial solute concentration, the freezing process occurs over a very broad range of temperature (up to $\Delta \textrm{U} \sim \SI{3}{\K}$ for $[\textrm{KCl}]=0.37~\textrm{M}$). The effect is again almost two orders of magnitude larger than for solute-free experiments, and therefore cannot be explained by freezing kinetics.

These observations can be explained by the accumulation of solute in the porous matrix throughout the freezing process. The solute concentration is initially the same in every pore. As freezing proceeds, bigger pores freeze first---due to the Gibbs-Thomson effect---and expel the solute they contained (Fig.~\ref{fig:Confinement}I). The local solute concentration increases thus as additional pores are freezing. Hence, solute and confinement have  a synergistic effects: confinement destabilizes the freezing front and allows solute to be expelled, enlarging the mushy layer and delaying the completion of freezing.

We can estimate the local $[\textrm{KCl}]$ concentration from Eq.~\ref{eq:GT_Solute} and $\Delta \textrm{U} (\textrm{KCl})$, the temperature range over which the freezing front spreads (Fig.~\ref{fig:Solute}E). By removing the kinetic contribution (obtained from solute-free experiments) and assuming this residue to be solely due to solute effects, we compute $\Delta [\textrm{KCl}]$ the local increase of solute concentration upon freezing (Fig.~\ref{fig:Solute}F, see SOM). A three-fold increase of salt concentration is expected. The estimated maximum concentration reached in these experiments ($\sim 1~\textrm{M}$) is approaching the solubility limit ($3.8~\textrm{M}$) of KCl in water at $0^\circ$C. As further decrease of pore size will result in greater salt concentration, we can expect to eventually reach the salt crystallization, which can thus occur even for low initial concentrations.

Two effects are thus combined in the presence of salt. The colligative depression of the freezing point due to KCl lowers the temperature at which the first (largest) pores will freeze. Once freezing is initiated in the largest pores, the synergy between confinement and solute revealed here will further lower $T_{fc}$, the temperature at which freezing is completed (see SOM). These combined effects (Fig.~\ref{fig:Solute}G) are almost two orders of magnitude greater (up to $\sim \SI{4.2}{\K}$ for $[\textrm{KCl}]=0.37~\textrm{M}$) than for confinement only ($\sim \SI{0.15}{\K}$), and much greater than expected from a simple combination of both effects.


Because of the morphology of the mushy layer in the presence of salts, with isolated liquid pockets remaining (Fig.~S3), there is no easy pathway for the salt to diffuse away from these isolated concentrated regions, except for premelted films around the particles. We can therefore expect the high salt concentration reached locally to remain. This has important consequences in soils physics and plant life, where salinity effects on plants can be critical~\cite{Parida2005}, and can lead to pitting corrosion in materials such as reinforced concrete~\cite{Bertolini2013}.

The mushy layers spread over $\sim \SI{100}{\micro\m}$, even for low solute concentration. Temperature gradients in soils are much lower ($\sim \SI{10}{\K\per\m}$) than in our experiments, hence we can expect the mushy layer to reach macro-scales ($\lambda_{soil} \sim \SI{10}{\centi\m}$) in these conditions. This observation is particularly important as it could explain the frozen fringe observed between ice lenses in soils~\cite{Peppin2011}.

Salts do more than just depress the freezing point, and additional phenomena probably need to be taken into account. For instance, it was just suggested~\cite{Zhou2019} that the rejection of salts ions in confined environment during freezing may results in enormous disjoining pressures at the pore scale, which could explain the freezing-induced damages of construction materials such as concrete. Our results provides a practical and theoretical framework to understand and predict the salt concentration due to freezing at the pore scale and expand this analysis.

Overall, this study shows that solutes such as salts must be taken into account to explain the variety of freezing patterns observed experimentally--variety that we still struggle to explain today~\cite{Wettlaufer2019}. The synergy between salt and confinement is explained by the thermodynamics of freezing and solute rejection at the pore scale, and explain how small amounts of salts can have a large effect in freezing heterogeneous media such as soils or construction materials. This provides a general framework to describe not only the amount and spatial distribution of residual water in frozen media, but also the progressive salt concentration, which is known to greatly impact the cryotolerance properties of construction materials and many organisms. We expect the magnitude of these effects to be enhanced by reducing the pore size down to nm scale or smaller.

\bibliography{Freezing_under_Confinment.bib}
\bibliographystyle{Science}

\textbf{Acknowledgements}
We thank Benoit Coasne for the discussion on these results and analyses. \textbf{Funding}: The research leading to these results has received funding from the European Research Council under the European Community’s Seventh Framework Programme (FP7/2007-2013) Grant Agreement no. 278004. T.L.'s internship was funded by Saint-Gobain. \textbf{Author contributions}: S.D. designed and supervised the project, S.D. and F.G. designed the experiments, D.D. designed and built the cooling stage, F.G. and T.L. carried out the confocal microscopy, F.G. wrote the code to analyze the data, S.D., F.G. and T.L. analyzed the data. All authors discussed the results and implications. F.G. and S.D. wrote the manuscript. \textbf{Competing interest}: The authors declare no competing interest. \textbf{Data and materials availability}: Additional data and code related to this paper are available on Figshare at [link to be added].

\textbf{Supplementary materials}
Detailed methods, Figures S1--S3, Movie S1, References 24--27


\newpage


\end {document}